# Whole-Genome Sequence of the *Trypoxylus dichotomus* Japanese rhinoceros beetle

Norichika Ogata
*Nihon BioData Corporation, 3-2-1 Sakado, Takatsu-ku, Kawasaki, Kanagawa 213-0012, Japan*
norichik@nbiodata.com

**Abstract.** The draft whole-genome sequence of the Japanese rhinoceros beetle, *Trypoxylus dichotomus* was obtained using long-read PacBio sequence technology. The final assembled genome consisted of 739 Mbp in 2,347 contigs, with 24.5× mean coverage and a G+C content of 35.99%.

Genomic research of the Japanese rhinoceros beetle, *Trypoxylus dichotomus* has been of interest since their large body size (~35g) and sexual dimorphizm in adult horn, and although several transcriptome studies had been conducted (1-3), the genome itself has not yet been deciphered. Here we sequenced and assembled the genome of the beetle. Last-instar larvae were harvested at Fuchu-shi, Tokyo, Japan and reared in leaf mold. Larval fat bodies were isolated. Total DNA was extracted from the fat bodies using the Qiagen Blood & Cell Culture DNA Maxi Kit (Qiagen, Gaithersburg, MD). Isopropanol and the eluted DNA mixture were dispensed to 1.5 mL tube at the step of centrifuging DNA for making pellet.

DNA shearing performed using Megaruptor® 3 (Diagenode, Liège, Belgium), targeting an average fragment size of 20 kb. The SMRTbell Express template preparation kit 2.0 (Pacific Biosciences, Menlo Park, CA) and the Barcoded Overhang Adapter Kit (Pacific Biosciences) were used to ligate hairpin adapters required for sequencing to the fragmented DNA. The library was size selected using the SageELF (Sage Science, Beverly, MA) and the 0.75% Agarose DNA ELF cassette with marker 75 (Sage Science). Sequence template was prepared by Sequel II Binding Kit 2.0 (Pacific Biosciences) and Sequel II DNA Internal Control Kit 1.0 (Pacific Biosciences). Sequencing was done on the Sequel II System (Pacific Biosciences) using Sequel II Sequencing Kit 2.0 (Pacific Biosciences) and Sequel SMRT Cell Oil (Pacific Biosciences). Data processing was performed using SMRT Link v8.0.0 (Pacific Biosciences). In total, 18,816,210,693 bp, 1,363,523 reads were generated as CCS reads.

Reads were trimmed, corrected, and assembled using the Hifiasm software version 0.3.0(4). The final assembled genome consisted of 739,405,423 bp in 2,347 contigs, with 24.5× mean coverage. The largest contig size was 36,507,117 bp. The N50 of the contigs was 7,929,129 bases. The G+C content of the beetle genome was 35.99%. Library preparation and sequencing were performed at the Takara Bio Inc. (Kusatsu, Shiga, Japan).

This whole-genome shotgun project has been deposited at GenBank under the accession no. BNES00000000.1. Raw reads were deposited at DDBJ SRA under the accession no. DRA010769.